\documentclass{aastex631}
\usepackage{graphicx}
\usepackage{float}
\usepackage[verbose]{wrapfig}
\usepackage{url}
\usepackage{natbib}
\usepackage{verbatim} 

\begin{document}
\title{Progenitors of Low Redshift Gamma-ray Burst}

\author{Vah\'e Petrosian}

\affiliation{Department of Physics  Stanford University, 382 Via Pueblo Mall, Stanford, CA 94305-4060}
\affiliation{Kavli Institute for Particle Astrophysics and Cosmology, Stanford University}
\affiliation{Also Department of Applied Physics, Stanford University}

\author{Maria G. Dainotti}

\affiliation{National Astronomical Observatory of Japan, Mitaka, Tokyo 181-8588, Japan}

\affiliation{The Graduate University for Advanced Studies, SOKENDAI, Kanagawa 240-0193, Japan}

\affiliation{Space Science Institute, Boulder CO,  80301, USA}

\begin{abstract}

Bimodal distribution of the observed duration of gamma-ray bursts (GRBs) has led to two distinct progenitors; compact star mergers, either two neutron stars (NSs) or a NS and a black hole (BH), for short GRBs (SGRBs), and so-called collapsars for long GRBs (LGRBs). It is therefore expected that formation rate (FR) of LGRBs should be similar to the cosmic star formation rate (SFR), while that of SGRBs to be delayed relative to the SFR. The localization of some LGRBs in and around the star forming regions of host galaxies and some SGRBs away form such regions support this expectation. Another distinct feature of SGRBs is their association with gravitational wave (GW) sources and kilonovae. However, several independent investigations of the  FRs of long and short bursts, using the  Efron-Petrosian  non-parametric method have shown a LGRB FR that is significantly larger than SFR at low redhift, and similar to the FR of SGRBs. In addition, recent discovery  of association of  a low redshift long GRB211211A  with a kilonova raises doubt about its collapsar origin. In this letter we review these results and show that low redshift LGRBs could also have compact star mergers as progenitor increasing the expected rate of the GW sources and kilonovae significanly.  
\end{abstract}

\section{Introduction}
\label{sec:intro}

Since the discovery of GRBs by Vela satellites several instruments on board  multiple satellites (CGRO,
Beppo-Sax, Fermi-GBM, Swift, Konus-Wind and others) have detected more than thousand GRBs, majority of 
which are classified as LGRBs and most of the rest as SGRBs. A good fraction of these GRBs have measured
spectroscopic or photometric redshifts, or redshifts based on host galaxies.   The bimodality of the 
distribution of  {\it observed} duration,  $T_{90}$, of GRBs was established  by the BATSE 
instrument of CGRO \citep{mazets,Kouveliotou1993ApJ...413L.101K}, dividing them into two classes of SGRBs and LGRBs separated at 
$T_{90}=2$ s. This feature has stayed robust and remained valid for the
subsamples with {\it rest frame} durations. However, the value of dividing $T_{90}$ is somewhat uncertain. A different classification in collapsar and not collapsar events of Swift GRBs by \cite{bromberg2013}, obtains a dividing $T_{90}\sim 0.8$s.
There are other features that separate the two classes. LGRBs  on average have softer spectra
than SGRBs, as measured either by their spectral hardness ratio or the value of
$E_p$, the photon energy
at the peak of the $\nu f(\nu)$ energy spectrum.%
\footnote{There have been some, but  not universally accepted, claims of 
existence of a third class of GRBs with intermediate values of duration and
spectral hardness. We will not address this possibility here.}  Localized LGRBs are often 
associated with star forming
galaxies and regions while SGRBs are found in older systems and away from star
forming regions (see,~review by \cite{berger2014}).
These dichotomies have led to two distinct progenitors.  LGRBs are believed to have been
produced by supernova explosions of massive Wolf-Rayet stars, so that their FR is 
expected to follow the cosmic SFR. SGRBs, on the other hand, are believed to have been 
produced during NS-NS or NS-BH mergers, thus their FR is expected to follow a delayed
form of the SFR. Such mergers are also expected to lead to kilonovae and  produce GWs.
Recent association of a  GW source with  SGRB 170817  \citep{troja2017,Abbott2017ApJ...848L..13A,Pian2017Natur.551...67P} has firmed this belief. 
The left panel of Figure \ref{fig:sfr} shows a recent determination of the cosmic SFR by \cite{MadauandDickinson2014ARA&A..52..415M},  and  examples of the delayed-SFR from \cite{Paul2018MNRAS.477.4275P} assuming a power law distribution of the delay time.

These dichotomies can be further tested by comparing the distributions of their intrinsic
characteristics, mainly the  luminosity function (LF),  and luminosity and FR  evolutions,
$L(Z)=L_0g(Z)$ and ${\dot \rho}(Z)$, all of which can be described by the
bi-variate distribution $\Psi(L,Z)$.%
\footnote{In what follows we use $Z=1+z$ as the redshift variable because for samples with large range of redshifts
it reflects the physical processes more directly and simplifies the equations. When necessary we assume 
a flat $\Lambda$CDM cosmological model with $\Omega_m=0.3$ and $H_0=70$ km s$^{-1}$ Mpc$^{-1}$.} 
In particular, a clearer picture of FRs of GRBs can provide a better estimation of rate of occurrence 
of  low mass merger GW sources, and kilonovae, especially at low redshifts that are within the  reach of current GW detectors. 

The main goal of this letter is to determine the FR of GRBs, ${\dot \rho}(Z)$, at low redshifts, $Z<3$. 
In the next section we describe the method we use to this end, and in \S 3 we review results obtained 
for FR of GRBs using this method, and present evidence which indicates that low redshift LGRBs could also
have compact star mergers as progenitors, which would increase expected rate of GW detection. A brief summary and conclusions are presented in \S 4.

\section{The Basic Problem and Procedures}
\label{sec:methods}

To achieve the above mentioned goal we need an accurate description of the general LF,  $\Psi(L,Z)$.
This of course requires samples of GRBs with known redshifts,%
\footnote{Among ($>1500$) {\it Swift} LGRBs there are several hundreds with
redshifts that can be used for this purpose.}
and a knowledge of all observational selection effects, which truncate the data and introduce bias. 
The attempts to correct for this bias, known as the {\it Eddington or Malmquist Bias,} has had a long
history. Early attempts, including a majority of current ones dealing with GRBs,  
have used parametric forward fitting
methods, whereby a set of 
functional forms are fit to the data to determine the ``best fit values" of the
parameters, using $\chi^2$ or maximum likelihood algorithms. These methods  assume parametric forms for 
several  functions;  spectrum, light curves, $g(Z)$ and ${\dot \rho}(Z)$,  each with three or more
parameters, raising serious questions about the {\it uniqueness} of the results.
They also require binning and thus large samples, which is not always the case, especially for SGRBs.
There are several  {\it non-parametric, non-binning methods} (e.g.~the $V/V_{\rm
max}$, \cite{Schmidt1968ApJ...151..393S};  $C^-$, \cite{Lynden-Bell1971MNRAS.155...95L}) that are more appropriate for small samples.
However, a major
drawback of these methods is the {\it ad hoc} assumption of {\it independent or uncorrelated
variables}, which for extragalactic sources means no luminosity evolution, $g(Z)=$const. 
For more details, see a review by \cite{Petrosian1992scma.conf..173P}. To overcome this shortcoming,
\cite{Efron1992,EfronandPetrosian1999astro.ph..8334E}, (hereafter called the EP method) developed new methods, whereby one first determines
whether $L$ and $Z$ are correlated or not. If correlated, then it introduces a new variable 
$L_0\equiv L/g(Z)$ and finds the luminosity evolution function, $g(Z)$, that yields an uncorrelated $L_0$ and $Z$, whose 
distributions, LF $\psi(L_0)$ and FR ${\dot \rho}(Z)$, can be obtained using,  e.g.~the Lynden-Bell $C^-$ 
method. {\it The advantage of this combined EP-L method is
that the two evolutions  are obtained directly from the data,
non-parametrically
and without binning}.%
\footnote{A third important aspect is that with this method one can easily combine data from different instruments with different spectral responses and more complicated selection criteria than the commonly available single waveband thresholds.}
In the past, these methods have been proven to be  very useful for studies of GRBs \citep{Lloyd1999,Lloyd2002,KocevskiandLiang2006ApJ...642..371K,Yonetoku2004,Dainotti2013b,dainotti2015b,Dainotti2017a,Dainotti2017ApJ...848...88D,Dainotti2020a,Dainotti2021ApJ...914L..40D,Dainotti2021Galax...9...95D,Dainotti2022MNRAS.514.1828D}, and AGNs \citep{Maloney1999ApJ...518...32M,Singal2011ApJ...743..104S, Singal2016ApJ...831...60S, Singal2019ApJ...877...63S, Singal2022ApJ...935L..19P,DainottiQuasar2022ApJ, Singal2022ApJ...932..111S, 2023ApJS..264...46L}. 

\section{Results on Formation Rates}
\label{sec:FRs}

In this section we present results from more recent works on the FR of  GRBs  using  the non-parametric  EP-L method. As emphasized above this method requires a sample with measured redshift and well defined observational selection effects; samples referred to as ``complete". Most instruments have well defined  $\gamma$-ray peak  flux threshold and hence provide complete samples. However, securing a redshift is a complicated process and includes several other observational selection effects that are not as easily quantified as the peak flux limit. {\it Swift} has been most successful in attempting to quantify these effects. After the $\gamma$-ray trigger the on board X-ray instrument attempts  a more accurate  localization which is more likely for brighter sources. This introduces addition selection effect based on the X-ray flux limit. Then, the optical-UV telescope attempts to produces a spectrum, and thus possibly redshift, and   to obtain a higher resolution localization that is sent to ground based telescopes for identification of a host galaxy and further redshift measurement. This third step is more complex and hard to quantify. 

\subsection{LGRBs}
\label{sec:lgrb}

\begin{figure}
\centering
\includegraphics[width=0.45\textwidth]{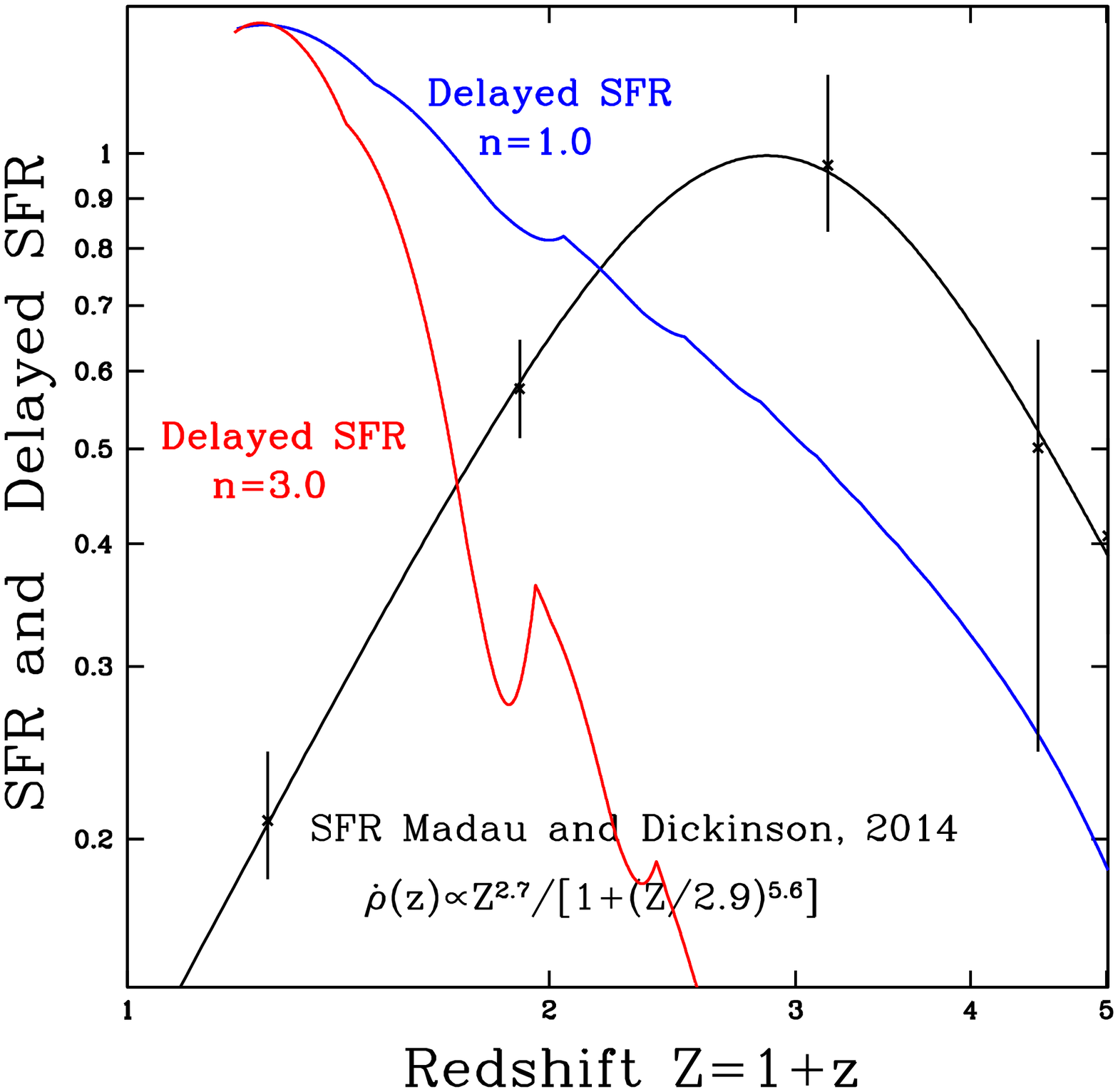}
\includegraphics[width=0.45\textwidth]{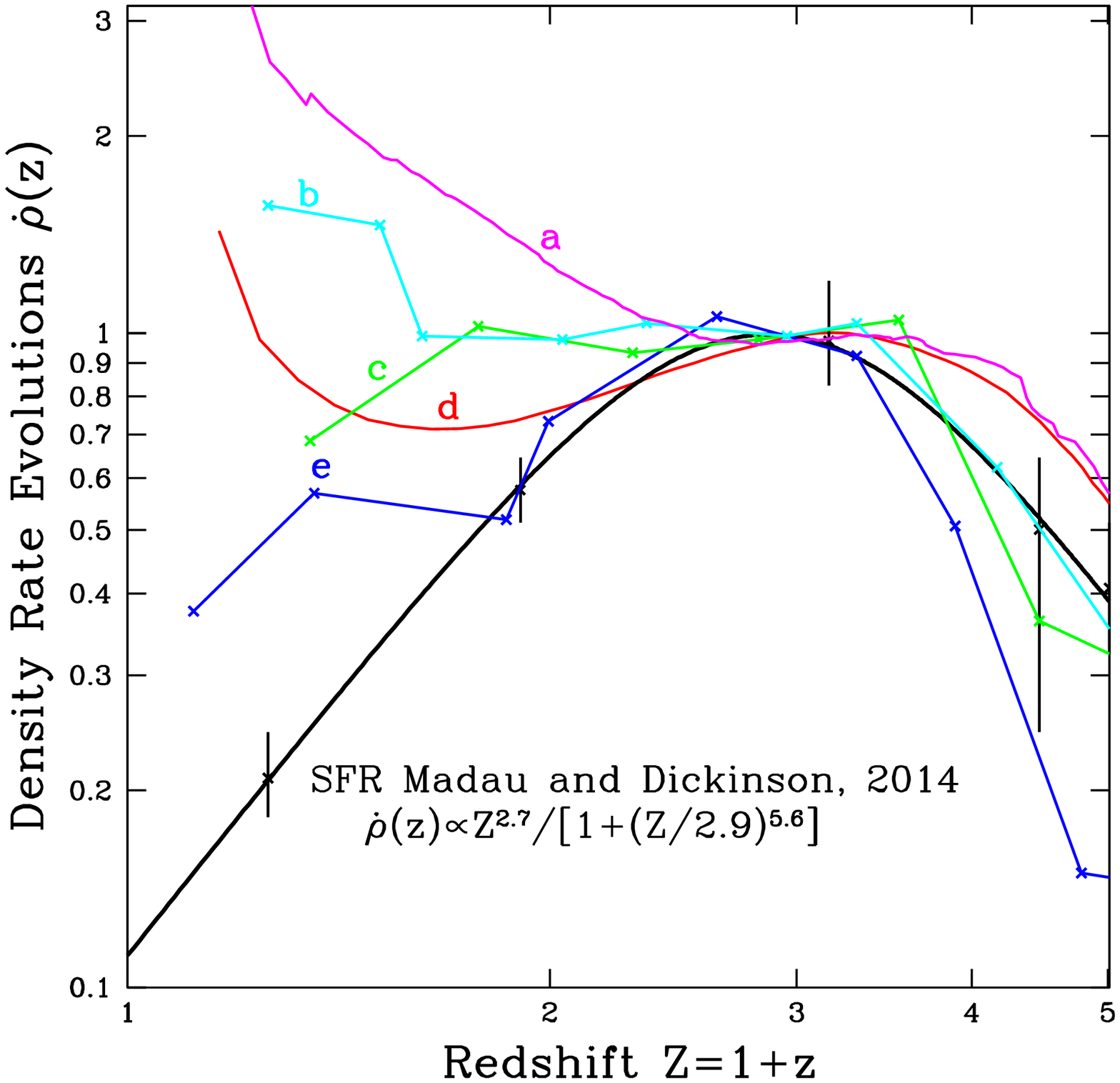}
\caption{{\it Left:} Cosmic SFR  from most comprehensive compilation  by \cite{MadauandDickinson2014ARA&A..52..415M} and two examples of Delayed-SFR with power law distribution of delay times, and  indexes 1.0 and 3.0; from \cite{Paul2018MNRAS.477.4275P} {\it Right:}  Results of 5 independent determination 
of LGRB FR, of somewhat different samples,  using the EP-L method, all normalized at the peak value of the observed SFR (black solid line with selected error bars): From top (a)-mganta \cite{Lloyd-Ronning2019MNRAS.488.5823L}; (b)-cyan \cite{Yu2015ApJS..218...13Y}; (c)-green \cite{Tsvetkova2017ApJ...850..161T} (d)-red \cite{Petrosian2015}; (e)-blue \cite{Pescalli2015}.  With different degrees, all are showing higher LGRB FR than  the SFR at low redshift (see text for more detailed comparison of these works).}
\label{fig:sfr}
\end{figure}

The right panel of Figure \ref{fig:sfr} shows five  results from five independent analyses, four of which use different samples of {\it Swift} LGRB data and one (in green) using  Konus-{\it WIND}  LGRB data \citep{Tsvetkova2017ApJ...850..161T}. It should be emphasized that none of the samples used in these works are  strictly speaking complete.  \cite{Yu2015ApJS..218...13Y} and \cite{Tsvetkova2017ApJ...850..161T} use the nominal $\gamma$-ray peak flux threshold of respective instruments. \cite{Lloyd-Ronning2019MNRAS.488.5823L} use fluence and energy $\epsilon_{\rm iso}$, instead of peak flux and luminosity, and several observed correlations. \cite{Pescalli2015} use a higher $\gamma$-ray threshold and obtain a sample with 80\% redshift completion. \cite{Petrosian2015}  use both the $\gamma$-ray and X-ray thresholds (based on X-ray flux compilation by \citet{nysewander09}), which yields significantly different FR histories than using the $\gamma$-ray threshold alone, especially at mid range redshifts. To account for more unknown optical selection bias \cite{Petrosian2015} use a $\gamma$-ray threshold 10 times larger than the nominal {\it Swift} threshold. As evident all of these estimates of FR of LGRBs deviate significantly, though at different degrees, from the cosmic SFR at $Z<3$ or $z<2$. It should be noted that \cite{Pescalli2015} compare their result with an earlier determination of the cosmic SFR which is flatter at low redshift than the more recent \cite{MadauandDickinson2014ARA&A..52..415M} one shown above, and conclude concordance with SFR. However, as shown in Figure \ref{fig:sfr}, their estimate, though to lesser degree than all others, shows a significant (about factor of 3; $\sim 5\sigma$) deviation at $Z<1.5$.%
\footnote{It should be noted that all 5 papers find a 2 to 3 $\sigma$ correlation between luminosity and redshift, indicating presence of  luminosity evolution with $g(Z)=Z^\alpha$ ($2<\alpha<3$). The FR evolution presented above corrects for this correlation.}

On the left panel of Figure \ref{fig:lowZ} we compare  the {\it average} value  of the logarithms of the above 5 FR estimates with the cosmic SFR.%
\footnote{Average of $\log({\dot \rho}(Z))$ is more appropriate. Straight averaging of the FRs would be dominated by the largest estimate of \cite{Lloyd-Ronning2019MNRAS.488.5823L}.}  
This average, not too different from the estimate by \cite{Petrosian2015}, which includes both gamma-ray and X-ray observational biases, clearly deviates from the SFR by more than  $10\sigma$'s at the lowest redshift bin.  Thus, we are  compelled to suggest the presence of two  LGRB components; one following the cosmic SFR and another low redshift component shown by the solid and dashed blue curves obtained from subtracting the first component from the observed average  and Petrosian et al. FR redshift distribution, respectively (again normalized at the peak of the cosmic SFR). This second FR component has an almost identical shape to that of  SGRB FR, shown by the magenta curve (described in the next section), and is  similar to the delayed SFR rates shown in the left panel Figure \ref{fig:sfr}.

\subsection{SGRBs}
\label{sec:sgrb}

\begin{figure}
\centering
\includegraphics[width=0.495\textwidth]{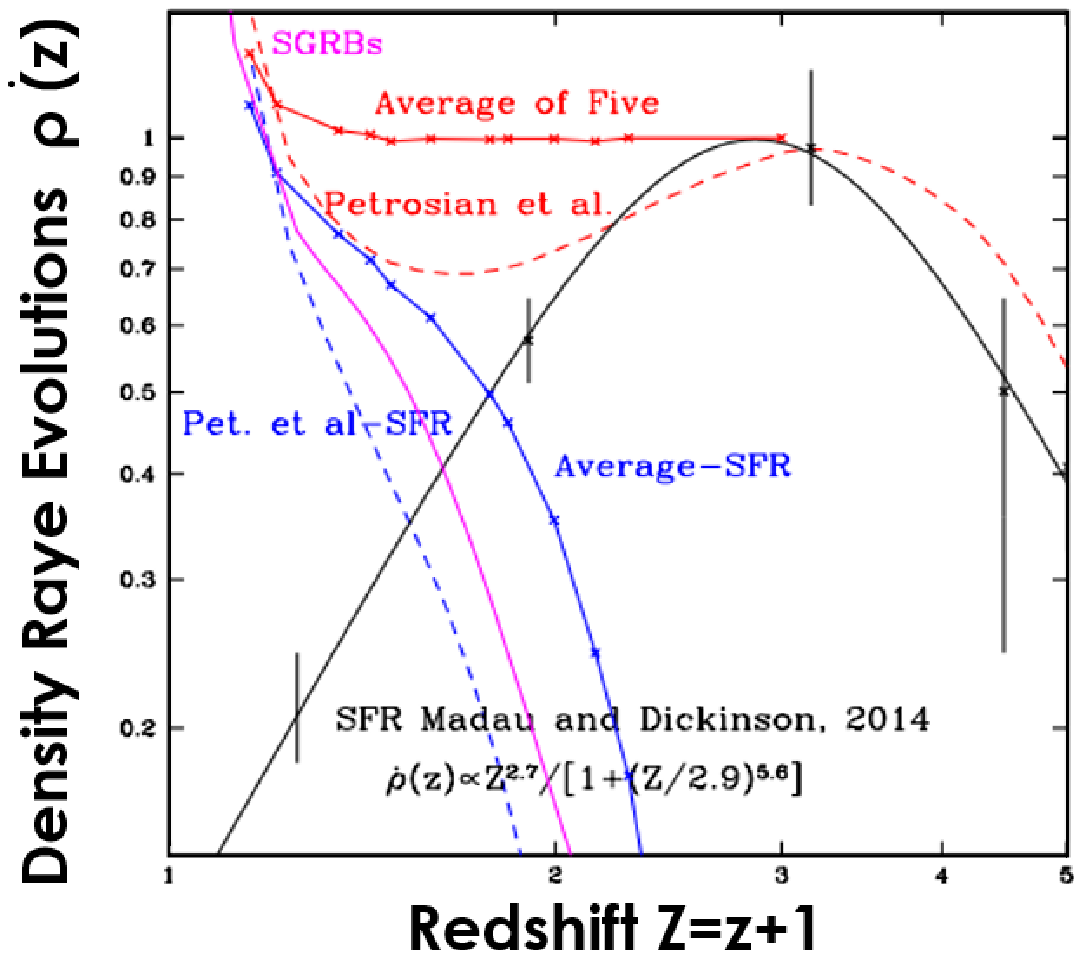}
\includegraphics[width=0.45\textwidth]{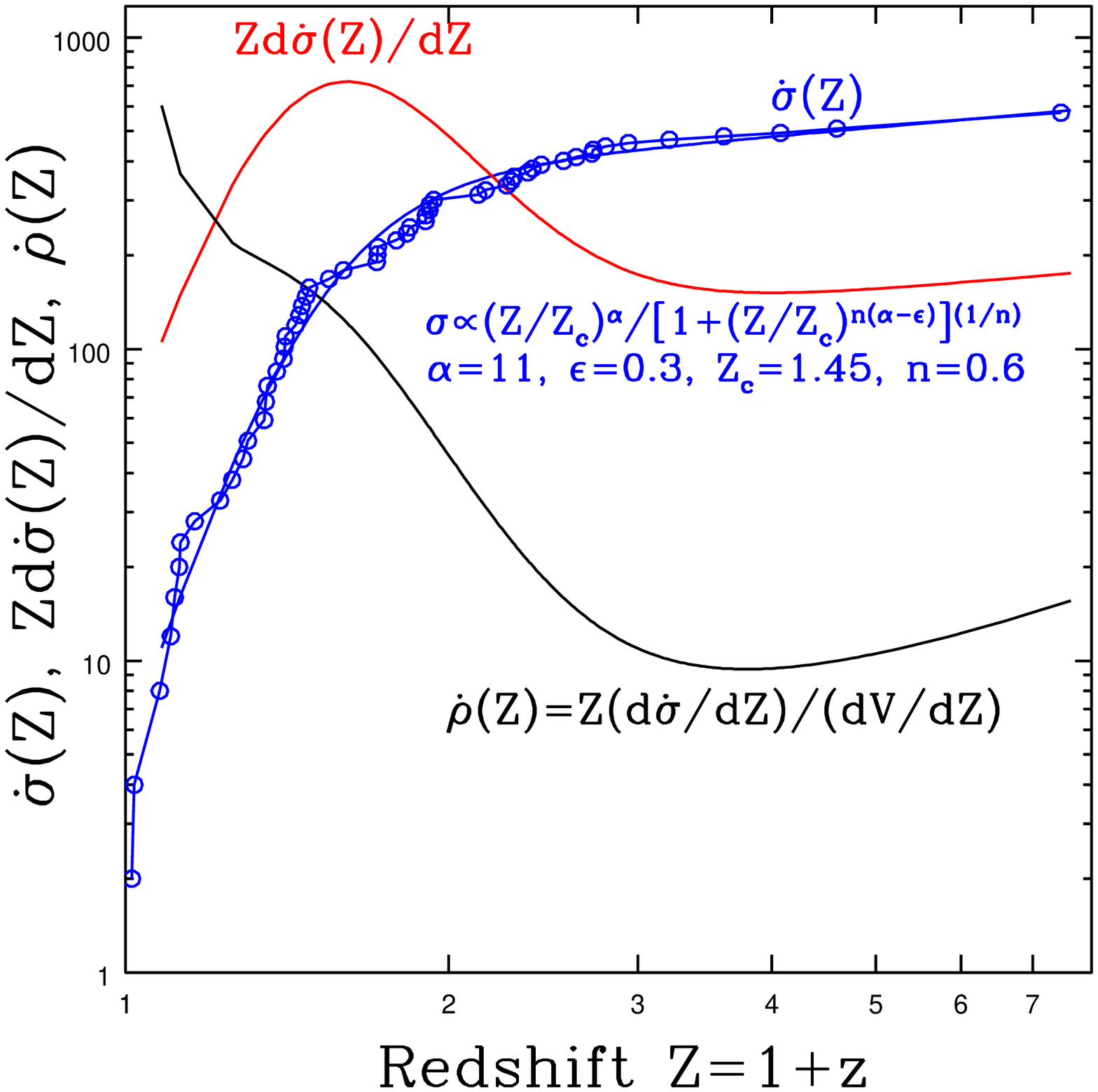}
\caption{{\it Left:} Same as Figure \ref{fig:sfr} left  but showing the average value of 5 independent estimates plus Petrosian et al. results in red. The blue lines show the two low  redsiuft components obtained bt subtracting the cosmic SFR from the observed ones. The magenta is the FR of SGRBs described in the next section.
{\it Right:} FR of SGRBs using most recent {\it Swift} sample of SGRBs with known redshift, using the procedures  described in \cite{Dainotti2021ApJ...914L..40D}. Here $\sigma(z)$ is the cumulative $(>Z)$ redshift distribution corrected for observational selection effects (blue dots), with arbitrary normalization. Blue curve is a broken power law fit (with indicated form anf fit parameters),  from which we obtain the FR, ${\dot \rho}(Z)$ analytically as described in the text.}
\label{fig:lowZ}
\end{figure}

The discoveries of gravitational wave sources  and the associations of some with SGRBs and kilonovae, mentioned above, has made  the
determination of the FR of SGRBs a
critical issue, and has led to increased activity on this front 
(e.g.~ \cite{Wanderman2015MNRAS.448.3026W,Ghirlanda2016A&A...594A..84G,Abbott2017ApJ...848L..13A,Paul2018MNRAS.477.4275P, ZhangandWang2018ApJ...852....1Z}). These are all based on 
parametric forward fitting method described above. More recently, \cite{Dainotti2021ApJ...914L..40D} have used the non-parametric EP-L method, for determining the FR od SGRBs, which we now described 
briefly. There are in general smaller numbers of SGRBs than LGRBs
and even smaller samples with known redshift. The size of the sample
is important. For larger sample one can obtain more reliable and 
detailed result no matter what method is used. But as emphasized 
above the observational selection that affect ones ability to 
determine the redshift is even more critical. The most important 
selection criteria is having a complete sample with well defined 
$\gamma$-ray peak flux threshold. In absence of X-ray observations and
due to small numbers, the more complete and conservative method used
by \cite{Petrosian2015} cannot be used for SGRBs. Instead, 
\cite{Dainotti2021ApJ...914L..40D} use a method of finding the lowest possible flux threshold, hence the largest sample, which statistically  indicates that the sample with redshift was drawn from the complete parent sample including all SGRBs. We have repeated this procedure with a slightly larger  SGRB sample. The results from this work are presented in the right panel of Figure \ref{fig:lowZ}. In general, the EP-L method yields  cumulative distributions of the variables, in this case the rate, ${\dot \sigma}(Z)$, of sources with redshift $>Z$, related to the differential rate as 
\begin{equation}
\label{sigma}
{\dot \sigma}(Z)=\int_0^Z Z'^{-1}{\dot \rho}(Z')dV/dZ', 
\end{equation}
where $V(Z)$  is the co-moving volume up to redshift $Z$.

The result are shown by the blue points and are fit by a broken power law,  parameters of which are given in the figure. The black line shows the RE obtained from differentiation of equation (\ref{sigma}) and the analytic form for ${\dot \sigma}(Z)$. This RE is shown by the magenta curve in the left panel normalized to the others at the lowest  redshift. As evident it has a form very similar to that of the low redshift component of LGRB FR.

Thus, it is reasonable to conclude that these result provide a strong evidence that both long and short low redshift GRBs have a common progenitor. The  fact that these FRs are similar to delayed SFR lead us to also conclude that mergers of compact stars (NSs and BHs) are the progenitor of all low redshift GRBs.
This conjecture has received strong support from a new observation showing coincidence in time and space of a  optical/NIR kilonova with a low redshift (z=0.076) LGRB (GRB211211A; \cite{Rastinejad2022Natur.612..223R}). 
Fermi-LAT observations show a transient (duration 20 ks) starting about 1 ks after the trigger that is interpreted to be due to inverse Compton scattering of kilonova photons by the GRB jet accelerated electrons \citep{Mei2022Natur.612..236M}. These  evidences complement the results presented here based on population studies.

\section{Summary and Conclusions}

In this letter we have explored how the FRs of long and short GRBs compare with cosmic SFR. The general expectation is that LGRBs produced by collapsars (supernova explosion of massive Woolfe-Rayet stars) should follow the SFR, while FR of SGRBs, produced by merger of compact stars (NSs and BHs),  should be delayed relative to SFR (Fig.~\ref{fig:sfr}, left). 

\begin{enumerate}

\item

We describe methods used for obtaining the FR history of extragalactic sources and the advantages of the non-parametric, non-binning Efron-Petrosian-LyndenBell method, which obtains the cosmological distributions and evolution of sources, yeilding the rate of luminosity evolution, the luminosity function, and the FR evolution, especially for small sample of sources with complex observational selection biases.

\item

We present results on five independent determination of the {\bf FR of LGRBs}, all using the EP-L method and showing significant deviations, though at different levels, from SFR at low redshifts (Fig.~\ref{fig:sfr}, right).

\item 

Taking the difference between the logarithmic average FR of 5 results and SFR, we show presence a distinct low redshift LGRB component that decreases rapidly with increasing redshift in a manner expected from delayed SFR (Fig.~\ref{fig:lowZ}, left).

\item 

We also present a new determination of {\bf FR of SGRBs}, again using EP-L method and data selection procedure described in \cite{Dainotti2021ApJ...914L..40D} (Fig.~\ref{fig:lowZ}, right). This FR is almost identical to that of the low reshift component of LGRBs (Fig.~\ref{fig:lowZ}, left).

\end{enumerate}

In summary, we have presented evidence for significant, and almost identical,  deviations of the FRs of both long and short GRBs from SFR at low redshifts, having a form expected from delayed SFR. The latter similarity is qualitative only but the assumption of power law distributions of delayed time used in \cite{Paul2018MNRAS.477.4275P} is questionable. For example, \cite{Wanderman2015MNRAS.448.3026W}, in addition to power law,  test log-normal distributions. There are other possibilities not explored yet.

{\bf Our main conclusion} is that the low redshift component of LGRBs, like the SGRBs, have compact merger progenitors. The evidence presented here in support of this comes from population study. An independent observation supporting this scenario comes from  recent discovery of association of a low redshift LGRBs with Kilonova discussed above.

For the future it will be useful to explore more quantitative comparison of the FR of low redshift GRBs with different delayed scenarios to determine the correct delay mechanism. Discovery of more association of low redshift LGRBs and Kilonovae will also be very helpful. 


\bibliography{sample}


\end{document}